\documentclass[
  reprint,
  amsmath,amssymb,
  aps,
  prd,
  superscriptaddress,
  nofootinbib
]{revtex4-2}

\usepackage{graphicx}
\usepackage{bm}
\usepackage{hyperref}

\begin{document}

\title{From Earth to Mars and Beyond: AI-Driven Cosmic-Ray Validation and Prediction with CorsikaAgent}

\author{Zhuo Chen}
\email{chenzhuo819@stu.pku.edu.cn}
\author{Zhizheng Zhao}
\author{Jinhong Shen}
\author{Leyun Gao}
\author{Cheng-en Liu}
\author{Qite Li}
\author{Zijian Wang}
\email{wangzijian@stu.pku.edu.cn}
\author{Qiang Li}
\email{qliphy0@pku.edu.cn}
\affiliation{School of Physics and State Key Laboratory of Nuclear Physics and Technology, Peking University, Beijing 100871, China}

\date{\today}

\begin{abstract}
We calculate the sea-level cosmic-ray muon flux with \textsc{CORSIKA} and
develop \textsc{CorsikaAgent}, an auditable large-language-model-based
system for configuring, executing, and validating air-shower simulations,
and for predicting particle fluxes in unexplored environments.  The
physics calculation uses proton primaries described by the solar-minimum
spectrum of Papini \textit{et al.}, with \textsc{UrQMD} at low interaction
energies and either \textsc{Sibyll}~2.3e or \textsc{QGSJET-III} at high
energies.  The predicted momentum spectra at approximately $0^\circ$ and
$75^\circ$ are compared with experimental measurements of the atmospheric
muon flux and with commonly used analytic parametrizations.  The two
interaction-model configurations give similar spectra and describe the
principal momentum and zenith-angle dependences of the data.  The agent
independently reconstructs the same observable through a different sampling
strategy and keeps every steering card, executable, seed, log, and
analysis script bound to the result it produced; its output is consistent
with the direct calculation at the level resolved by the present comparison.
Transferred to Mars, the workflow reproduces the \textsc{CORSIKA}~8
benchmark, and transferred to Venus it produces the first prediction of the
surface muon flux, an environment for which no measurement exists.  The agent
thus extends the workflow from auditable reproduction to first-principles
prediction: it autonomously reconfigures the simulation---atmosphere,
magnetic field, electric field---for new environments, while the proton-only
primary model, limited high-momentum statistics, and an incomplete
systematic-uncertainty treatment bound the physical precision that can
currently be claimed.
\end{abstract}

\maketitle

\section{Introduction}\label{sec:intro}

Muons are the dominant charged particles at sea level.  Their energy and
angular distributions reflect the primary cosmic-ray spectrum, the density
profile of the atmosphere, and the competition between interaction and decay
in the hadronic cascade~\cite{PDG2022,GaisserEngelResconi2016}.  Measurements
of the atmospheric muon flux therefore provide a direct test of air-shower
calculations and an essential input to underground experiments and muon
imaging.

\textsc{CORSIKA} is a standard Monte Carlo program for simulating extensive
air showers~\cite{Heck1998,CORSIKA78010}.  It transports the secondary
particles produced by a primary cosmic ray and delegates hadronic collisions
to selectable interaction models.  This modularity makes it possible to test
the model dependence of an observable, but it also creates a practical
burden: the executable must be compiled with a consistent set of options,
steering cards must respect the configuration of that executable, large event
samples must be managed, and particle-level output must be normalized with the
same sampling distribution used at generation.  Each step can produce a
plausible-looking but physically misnormalized result.

Atmospheric muon spectra have been measured over a wide momentum range.  The
L3+C analysis provides a precise spectrum between 20 and
$3000\,\mathrm{GeV}/c$ over several zenith-angle intervals~\cite{Achard2004},
while Jokisch \textit{et al.} measured the spectrum at a zenith angle of
$75^\circ$ up to the TeV scale~\cite{Jokisch1979}.  These data have motivated
both full shower calculations~\cite{Wentz2003,Djemil2005} and compact flux parametrizations
~\cite{Gaisser1990,Tang2006,Bugaev1998,Reyna2006,Guan2015}.  Comparisons among
such descriptions also show that agreement depends on momentum, zenith angle,
and observation conditions such as altitude and latitude~\cite{Su2021}.  At larger depths, the
KM3NeT detectors have compared the measured atmospheric muon flux with
\textsc{CORSIKA} simulations using the Global Spline Fit mass-composition
model~\cite{KM3NeT2023}.

Tool-using large language models (LLMs) offer one way to automate the
configuration and bookkeeping required by simulation workflows.  The relevant
capability is not text generation by itself, but the ability to alternate
between planning, tool execution, inspection of outputs, and correction of a
failed or inconsistent step~\cite{Yao2023,Schick2023}.  Similar ideas have
been explored in other scientific settings~\cite{Boiko2023,Bran2024}.  For a
physics calculation, however, successful execution is not sufficient:
physical inputs, software versions, random seeds, normalization conventions,
and validation evidence must remain inspectable.

This work has two connected aims.  First, we calculate the proton-induced
sea-level muon flux with CORSIKA~7.8010 using two high-energy interaction
models and compare it with data and analytic references.  Second, we develop
\textsc{CorsikaAgent}, an LLM-based workflow that repeats the calculation with
a different sampling strategy and applies the same validation procedure to
modified environments.  The agent calculation functions as a procedural
cross-check rather than an independent measurement: sharing the direct
analysis's physical assumptions and reference data, it confirms correct
execution but offers no separate test of those assumptions.  It can
nonetheless predict: where no measurement exists, the workflow carries the
validated simulation method into unmeasured environments.

These planetary cases are not incidental extensions.  Beyond the Earth, the
secondary-particle environment produced by cosmic rays controls the radiation
dose at planetary surfaces, constrains the habitability of near-surface
environments, and offers a remote probe of surface and subsurface structure.
The production and transport of muons in the Martian atmosphere, for
instance, have been examined for their relevance to the planet's
paleo-environment and geology~\cite{Tanaka2007}, while cascade-equation
studies of the Venus radiation environment report dose rates but no surface
muon spectrum~\cite{Globus2021}.  No surface muon flux has been measured
beyond the Earth, so simulation is presently the only route to such
quantities for other planets.  The cases studied below therefore serve
distinct purposes: Mars tests the transfer of the workflow against an
established calculation, Venus delivers a first prediction where no
measurement exists, and the thunderstorm configuration exercises the
electric-field modification in an environment with experimental reference
data.  In each case the same auditable chain---configuration, execution,
validation, artifact retention---is preserved, so that every prediction
carries explicit provenance.

\section{CORSIKA simulation and flux reconstruction}\label{sec:simulation}

This section describes the direct calculation, which is performed without the
agent and serves as the reference for the agent-based reconstruction of
Sec.~\ref{sec:agent}.  The simulations use CORSIKA~7.8010~\cite{CORSIKA78010}.  Primary particles
are restricted to protons.  This choice isolates the simulation and workflow
comparison, leaving a complete prediction for the all-particle cosmic-ray
composition for future work.  The primary intensity is taken from the
solar-minimum parametrization of Papini \textit{et al.}~\cite{Papini1996}.
Low-energy hadronic interactions are described by \textsc{UrQMD}~1.3.1
~\cite{Bass1998,Bleicher1999} in all runs, which reproduces the measured
low-energy secondary spectra more accurately than the older \textsc{GHEISHA}
package~\cite{Bhadra2009}; high-energy interactions are
simulated with either \textsc{Sibyll}~2.3e~\cite{Riehn2020,CORSIKA78010} or
\textsc{QGSJET-III}~\cite{Ostapchenko2024a,Ostapchenko2024b}.  The model
versions were verified directly against the executables and data files used
for the production: the \textsc{Sibyll} binary reports version~2.3e, and the
\textsc{QGSJET-III} binary reports \textsc{QGSJET-III} together with its
\texttt{qgsdat-III} cross-section tables.

All simulations use a build of CORSIKA~7.8010~\cite{CORSIKA78010} with the
curved-atmosphere option enabled, and with the electric-field option compiled
in but set to zero.  The standard US atmosphere is used, with no external
atmosphere tables or seasonal variation, and the observation
level is sea level.  The geomagnetic field is included through the
\texttt{MAGNET} card; no geomagnetic cut-off is applied.  All simulations are
performed at sea level, and the reference fluxes are used as published,
without altitude correction or site matching.

For the differential
momentum spectra, protons are injected isotropically over the downward
hemisphere (zenith angles $0$--$89^\circ$), with primary energies sampled
uniformly over $1$--$10^5\,\mathrm{GeV}$ and weighted by the Papini
solar-minimum spectrum.  To keep usable statistics across the steeply falling
spectrum, the sample is split into two energy segments: $5\times10^7$ protons
between $1$ and $100\,\mathrm{GeV}$, and $5\times10^6$ protons between $100$
and $10^5\,\mathrm{GeV}$.  The
spectral comparisons select muons arriving near $0^\circ$ and $75^\circ$.

The normalization of Eq.~\eqref{eq:flux} follows from the sampling geometry.
The cosmic-ray primaries are taken to be isotropic at the top of the
atmosphere and are generated over the full downward hemisphere with flat
$\sin\theta\cos\theta$ sampling, a distribution fixed by the isotropy of the
incident cosmic-ray flux and the planar observation surface.  Let
$\Phi_p(E)$ be the differential primary intensity per unit energy, area,
solid angle, and time.  The rate of primaries
crossing a unit horizontal area is obtained by projecting the isotropic
intensity onto the horizontal plane and integrating over the upper hemisphere,
\begin{equation}
  \int_{\mathrm{hemisphere}} \Phi_p \cos\theta\, d\Omega
  = 2\pi \Phi_p \int_0^{\pi/2} \cos\theta \sin\theta\, d\theta
  = \pi \Phi_p ,
\end{equation}
which is the origin of the factor $\pi$ in Eq.~\eqref{eq:flux}.  The effective
injection area and the sea-level collection area are both the Earth's surface,
so the area factors cancel between the primary rate and the muon
normalization, and no explicit area enters the estimator.  The integrated
primary flux over the generated energy range is
\begin{equation}
  F_p = \int_{1\,\mathrm{GeV}}^{10^5\,\mathrm{GeV}} \Phi_p(E)\, dE
      = 0.209\ \mathrm{cm^{-2}\,s^{-1}\,sr^{-1}} ,
\end{equation}
for the solar-minimum Papini spectrum~\cite{Papini1996}.  The product $\pi F_p$
thus converts the simulated muon yield per primary into an absolute flux per
unit horizontal area.

For clarity, the normalization used for the binned direct sample is written
explicitly.  Let $\Delta N^{(w)}_{\mu,kj}$ be the Papini-weighted number of
muons in momentum bin $k$ of width $\Delta p_k$ and solid-angle bin $j$ of
size $\Delta\Omega_j$, and let $N$ be the number of generated primaries.
With $F_p$ denoting the primary proton intensity integrated over the generated
energy interval, the estimator used in the analysis is
\begin{equation}
 \left.\frac{d^2\Phi_\mu}{dp\,d\Omega}\right|_{kj}
 = \frac{\pi F_p}{N\,\Delta p_k\,\Delta\Omega_j\cos\theta_j}
   \Delta N^{(w)}_{\mu,kj}.
 \label{eq:flux}
\end{equation}
Equation~\eqref{eq:flux} is the differential muon flux per unit momentum and
per unit solid angle, with $\theta_j$ the zenith angle of the muon
direction. Primary energies are sampled uniformly, so each simulated shower is
reweighted to the physical Papini spectrum before the results are combined.

No detector response is folded into the calculation.  The comparison is
therefore made with published fluxes or parametrizations at the corresponding
observation level, rather than with raw detector counts.  Statistical
fluctuations arise from the finite shower samples.  Uncertainties in the
primary spectrum, atmosphere, detector corrections of the reference data, and
hadronic-model parameters are not propagated in the present study, and the
model comparisons that follow should be read with this gap in mind.

\section{Muon-flux results}\label{sec:results}

\begin{figure}[t]
\centering
\includegraphics[width=\columnwidth]{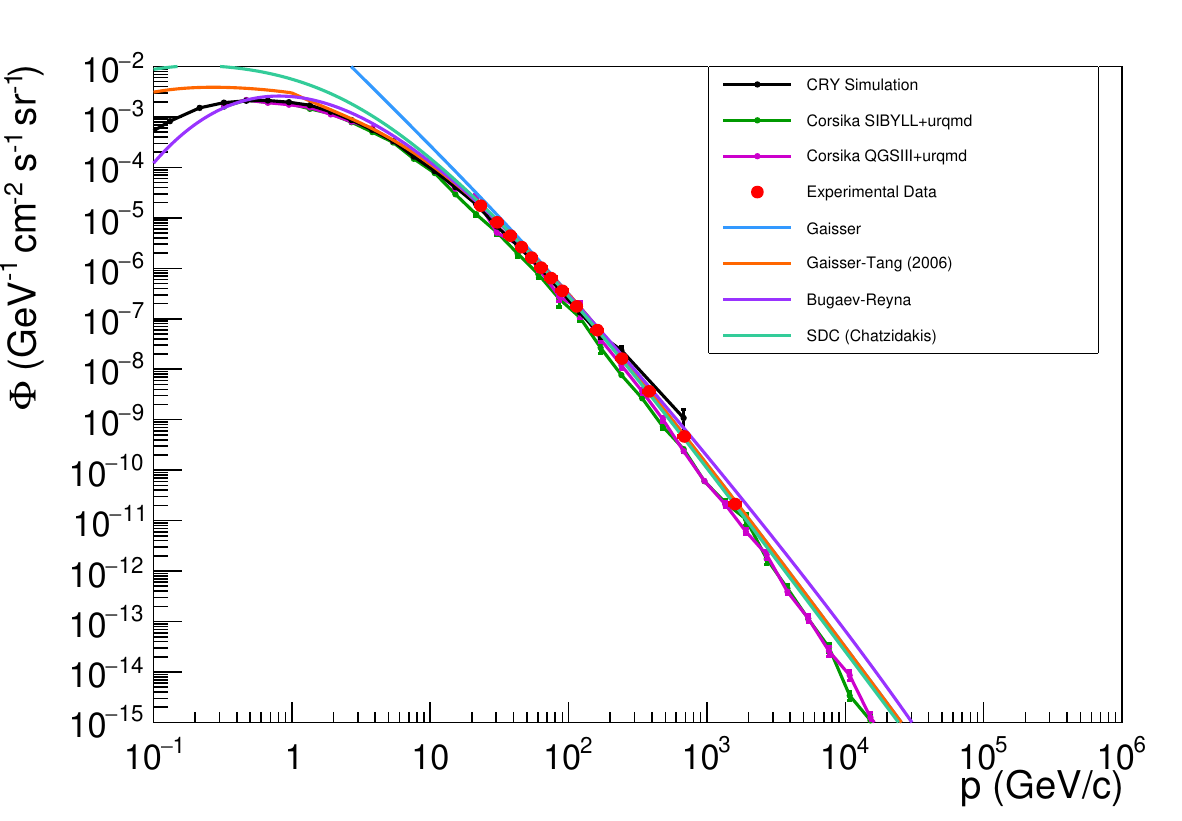}\\[0.8em]
\includegraphics[width=\columnwidth]{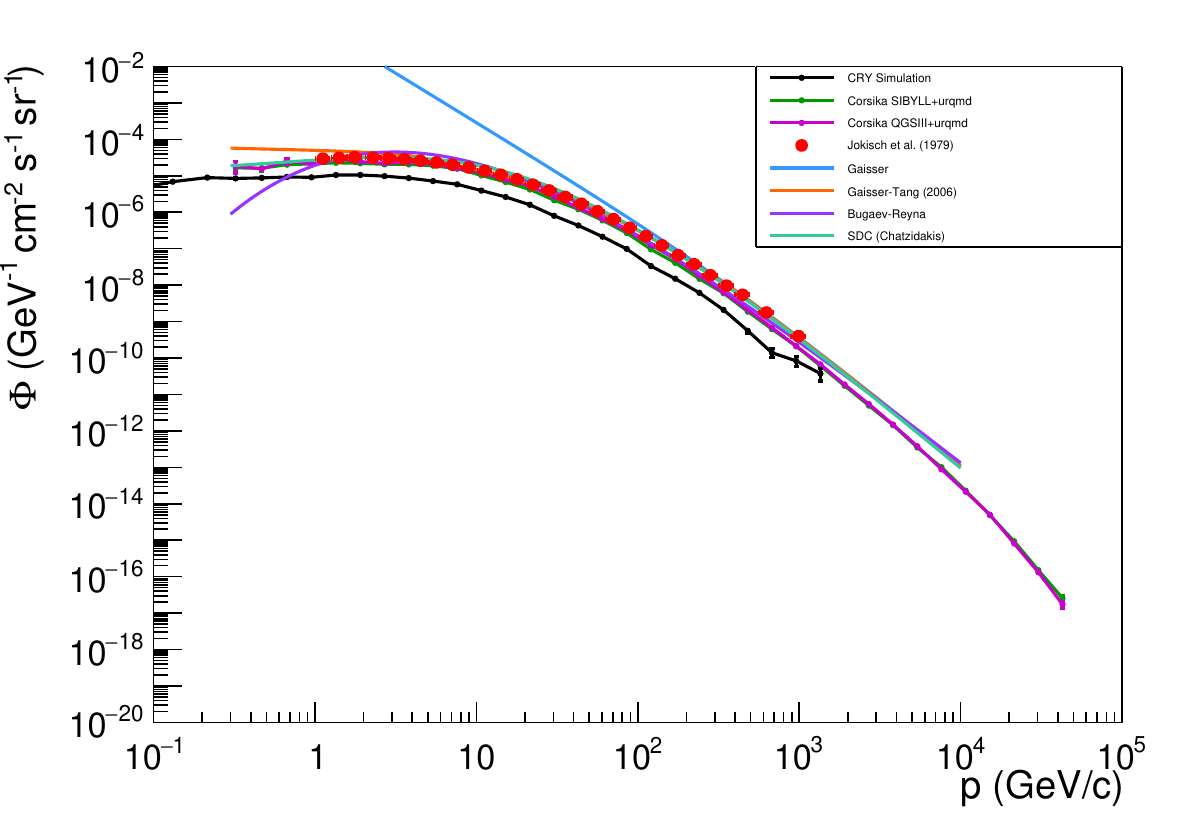}
\caption{Differential sea-level muon intensity versus momentum near
$\theta=0^\circ$ (top) and $\theta=75^\circ$ (bottom).  CORSIKA calculations
with \textsc{Sibyll}~2.3e or \textsc{QGSJET-III}, each coupled to
\textsc{UrQMD}, are compared with CRY~\cite{Hagmann2007}, analytic
parametrizations~\cite{Gaisser1990,Tang2006,Bugaev1998,Reyna2006,Smith1959,
Chatzidakis2015}, and experimental data.  The inclined data are from Jokisch
\textit{et al.}~\cite{Jokisch1979}.}
\label{fig:spectra}
\end{figure}

The differential spectra of the direct calculation are shown in
Fig.~\ref{fig:spectra}.  Near the
vertical direction the calculation reproduces the shape of the measured
spectrum, including the broad low-momentum maximum and the subsequent
power-law decrease, and the two interaction models give nearly identical
results on the scale of the plot.  At the lowest momenta the
\textsc{QGSJET-III} spectrum lies slightly above the \textsc{Sibyll}~2.3e one.
At $75^\circ$ the low-momentum flux is strongly depleted by the longer slant
path, and above a few hundred GeV the inclined spectrum
approaches and eventually exceeds the vertical one, in agreement with the
measured angular dependence and with the increasing probability that an
energetic parent meson decays before interacting along an inclined
trajectory~\cite{Gaisser1990,Bugaev1998}.

\begin{figure}[t]
\centering
\includegraphics[width=\columnwidth]{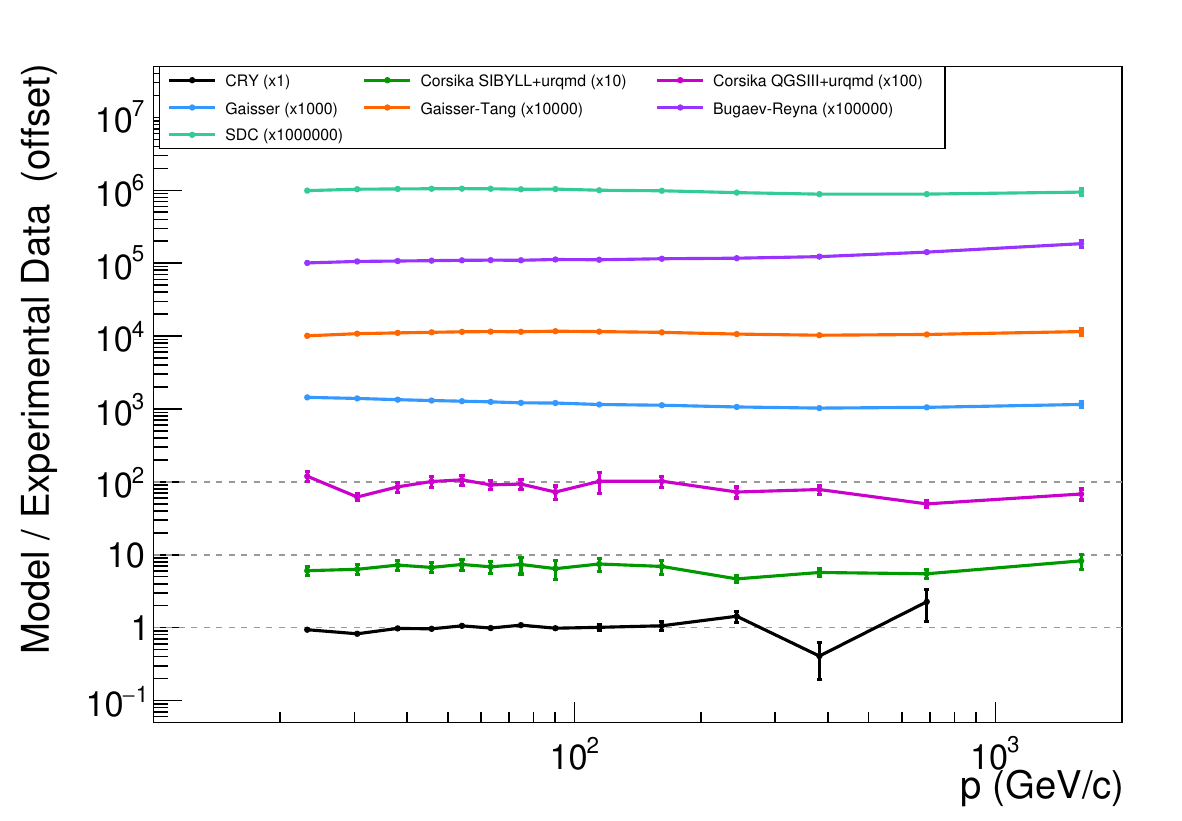}\\[0.8em]
\includegraphics[width=\columnwidth]{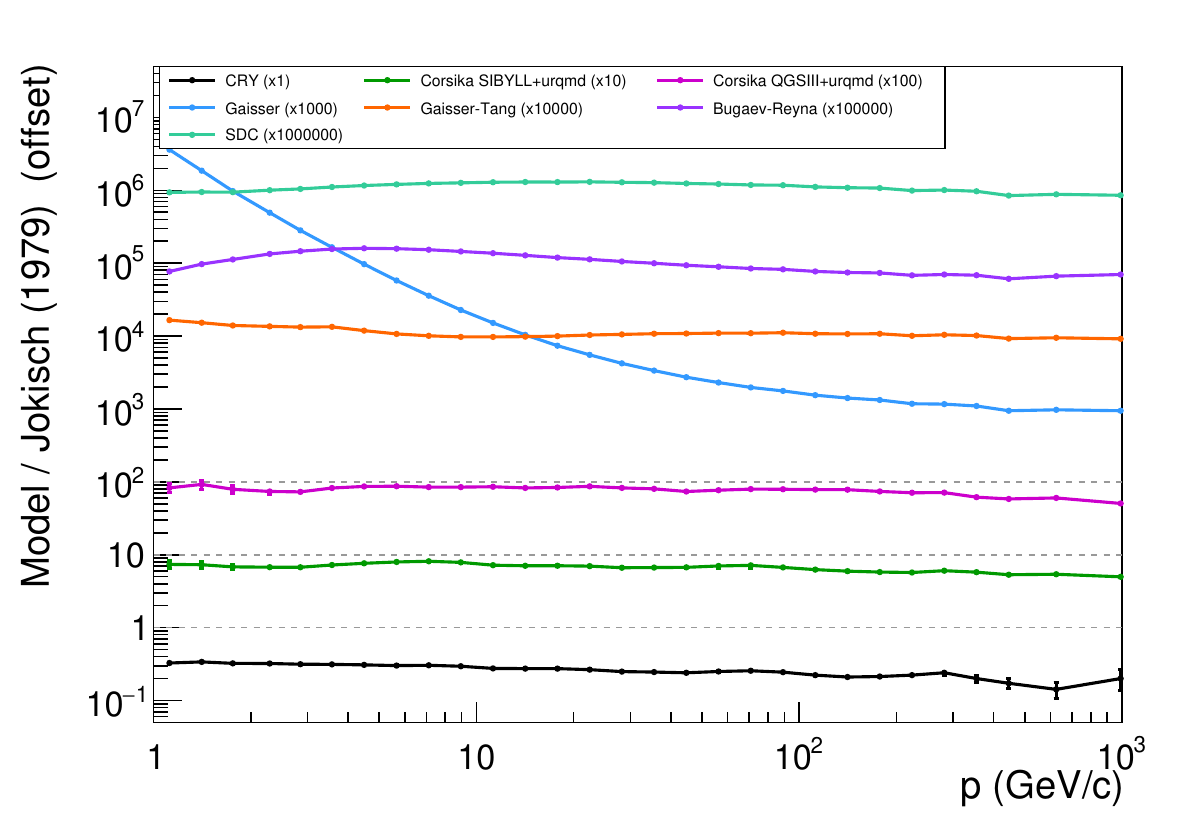}
\caption{Ratio of each prediction to the experimental reference near
$0^\circ$ (top) and to the Jokisch \textit{et al.} data at $75^\circ$
(bottom).  Successive curves are multiplied by powers of ten solely for
visibility, as specified in the legends.  The dashed horizontal guides must
be interpreted with the same offsets.}
\label{fig:ratios}
\end{figure}

The ratios in Fig.~\ref{fig:ratios} expose the differences that are
compressed on the logarithmic spectra.  The calculated fluxes generally lie
below the measurements over the displayed range, with a deficit that varies
with momentum and zenith angle.  Part of
this deficit is expected for a proton-only primary beam: a complete primary
composition would raise the predicted flux.  The comparison also separates the simulation
frameworks: at $0^\circ$ the CRY curve lies closest to the data, but at
$75^\circ$ it underestimates the flux by roughly a factor of three, whereas
both \textsc{CORSIKA} configurations track the data without such a
divergence.  The underestimation of CRY at large zenith angles is in line with the
comparison of Ref.~\cite{Su2021}; the present calculation, in contrast,
employs the curved-atmosphere option of \textsc{CORSIKA}
(\texttt{\_\_CURVED\_\_}) and therefore describes the inclined geometry
correctly.  Among the analytic parametrizations, Gaisser--Tang and
Bugaev--Reyna remain closest to the data, while the Gaisser form
overestimates the low-momentum flux, in agreement with earlier
comparisons~\cite{Su2021}.  A quantitative agreement level is not quoted, since many factors affect the
comparison and systematic uncertainties are not considered in this study.

These comparisons support two limited conclusions.  First, the normalization
procedure gives the correct overall scale without an additional fit to the
muon data.  Second, changing the high-energy interaction model between the two
options considered here has a smaller effect than the full spread among the
displayed references.  A precision prediction of the total atmospheric muon
flux would additionally require heavy primary nuclei and several sources of
systematic uncertainty that remain outside the present scope.

\section{Agent-based reconstruction and prediction}\label{sec:agent}

\textsc{CorsikaAgent} was developed to reconstruct a simulation from a
physics request while retaining an auditable chain from the request to the
final plot, and to extend the same workflow---atmosphere, magnetic field,
electric field---to new environments where it produces first-principles
predictions.  The implementation uses the open-source
\textsc{OpenCode} runtime~\cite{OpenCode} with the
\textsc{DeepSeek-V4-Flash} model~\cite{DeepSeekV4}.  The model is not treated
as a source of physics facts.  Its role is to select and execute tools, inspect
their outputs, and revise the next action subject to information stored in a
set of domain-specific prompts.

The agent's knowledge is organized into four curated prompts: (i)
\texttt{SKILL.md}, which defines the agent's workflow---planning,
configuration, execution, validation, and artifact retention; (ii)
\texttt{usage-guide.md}, a condensed reference of the \textsc{CORSIKA}
steering-card syntax and output formats; (iii)
\texttt{code-architecture.md}, a map of the relevant \textsc{FORTRAN} source
and build configuration; and (iv) \texttt{planetary-migration.md}, which
encodes the procedure, with authoritative parameters and sources, for
reconfiguring the atmosphere, gravity, and magnetic field of other planets,
as applied below to Mars and Venus.  These prompts are loaded only when the task
requires them.  Because they document how to modify the atmosphere, the
magnetic field, and the electric field---through steering cards or source
changes---the same agent reconfigures the simulation for environments far
from the terrestrial one, as exercised below for Mars, Venus, and a
thunderstorm, rather than being limited to reproducing the sea-level
calculation.  The executable, its \texttt{config.h}, steering card, random
seeds, standard output, and generated particle files are retained together.
The purpose is to bind a numerical result to the software configuration that
produced it.  A syntactically valid card is not accepted merely because the
run terminates: the selected interaction models and compile-time options are
checked against the installed binary and log output.

\begin{figure}[t]
\centering
\includegraphics[width=\columnwidth]{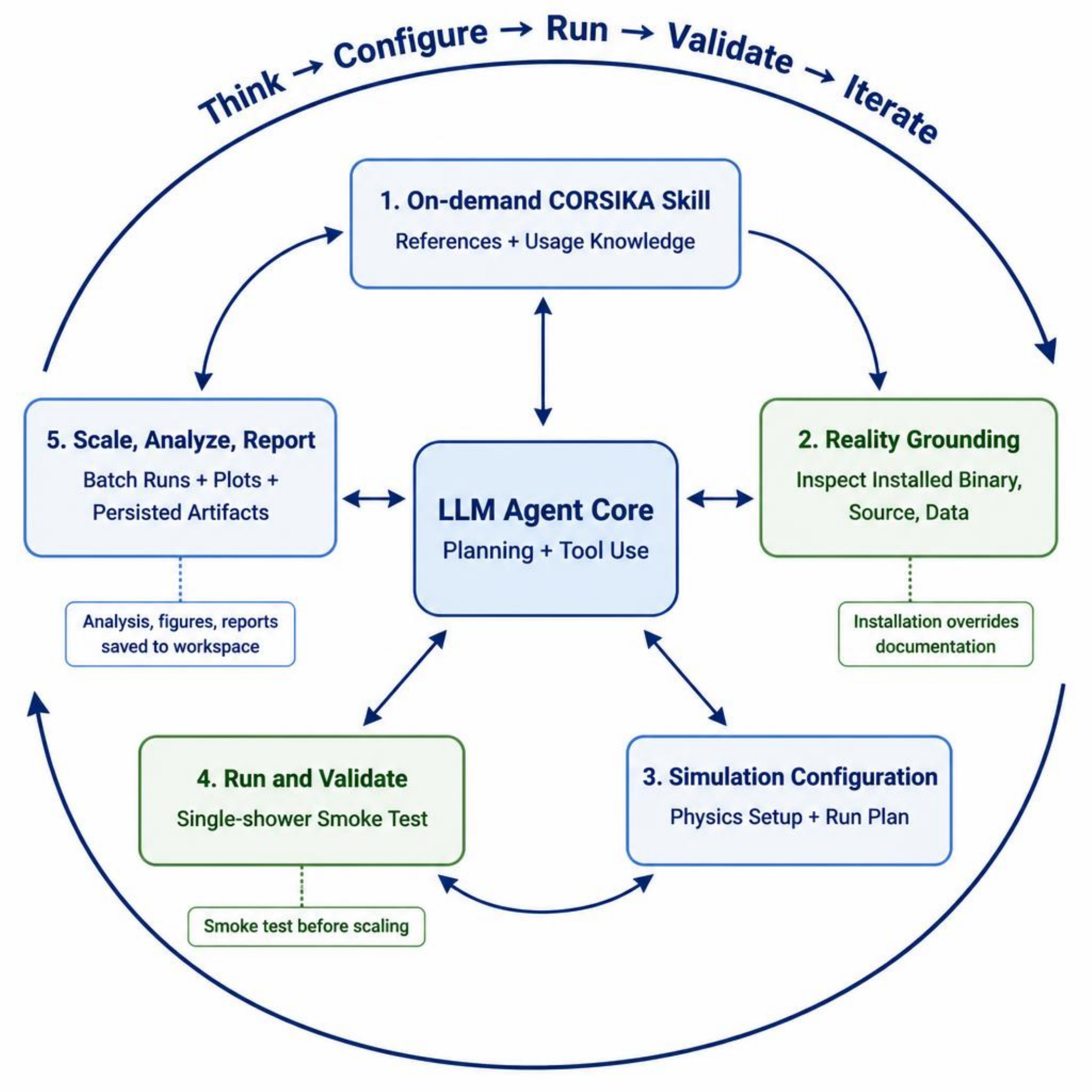}
\caption{Control loop used by \textsc{CorsikaAgent}.  The agent alternates
between planning, configuration, execution, validation, and revision.  Local
documentation supplies domain constraints, while the executable, build
configuration, smoke-test output, and production logs provide run-specific
evidence.}
\label{fig:agent}
\end{figure}

Figure~\ref{fig:agent} summarizes the control loop.  The agent first translates
the requested observable into a set of simulation configurations.  It then
checks the configuration against the local installation and executes a
single-shower or small-sample smoke test.  Particle identifiers, observation
level, energies, directions, and output-file structure are inspected before
the job is enlarged.  A source-level modification triggers a fresh
\texttt{coconut}/\texttt{make} cycle and another smoke test.  The steering
cards, seeds, logs, \texttt{DAT} files, analysis scripts, and plots of the
production runs are persisted.  Failed
checks return control to configuration or analysis rather than allowing the
workflow to proceed silently.

For the muon-flux cross-check, the agent deliberately uses a production design
different from the direct calculation.  Each primary-energy point is simulated
separately at fixed zenith angles of $0^\circ$ and $75^\circ$.  Concentrating
events at the requested directions avoids spending showers elsewhere in the
downward hemisphere.  The resulting energy-grid yields are weighted with the
same Papini proton spectrum and summed to obtain the momentum distribution.
Thus the two calculations share their physical inputs but differ in their
proposal distributions, steering cards, and aggregation paths.  Agreement is
therefore sensitive to errors in sampling weights and angular normalization,
though a physical assumption shared by both calculations would remain
invisible to this particular check.

The fixed-energy samples are generated on a grid of 21 primary energies,
uniformly spaced in $\log_{10}E$ with five points per decade between $10$ and
$10^5$~GeV; the number of showers per point increases towards lower energies.
The grid yields
are combined with midpoint weights in $\ln E$, and each muon-momentum bin is
populated directly from its grid point without further interpolation.  Since
the muon yield varies smoothly with primary energy, the residual bias of this
piecewise-constant treatment is of order $(\Delta\ln E)^2/24 \simeq 1\%$,
below the statistical uncertainties of the present samples.

\begin{figure}[t]
\centering
\includegraphics[width=\columnwidth]{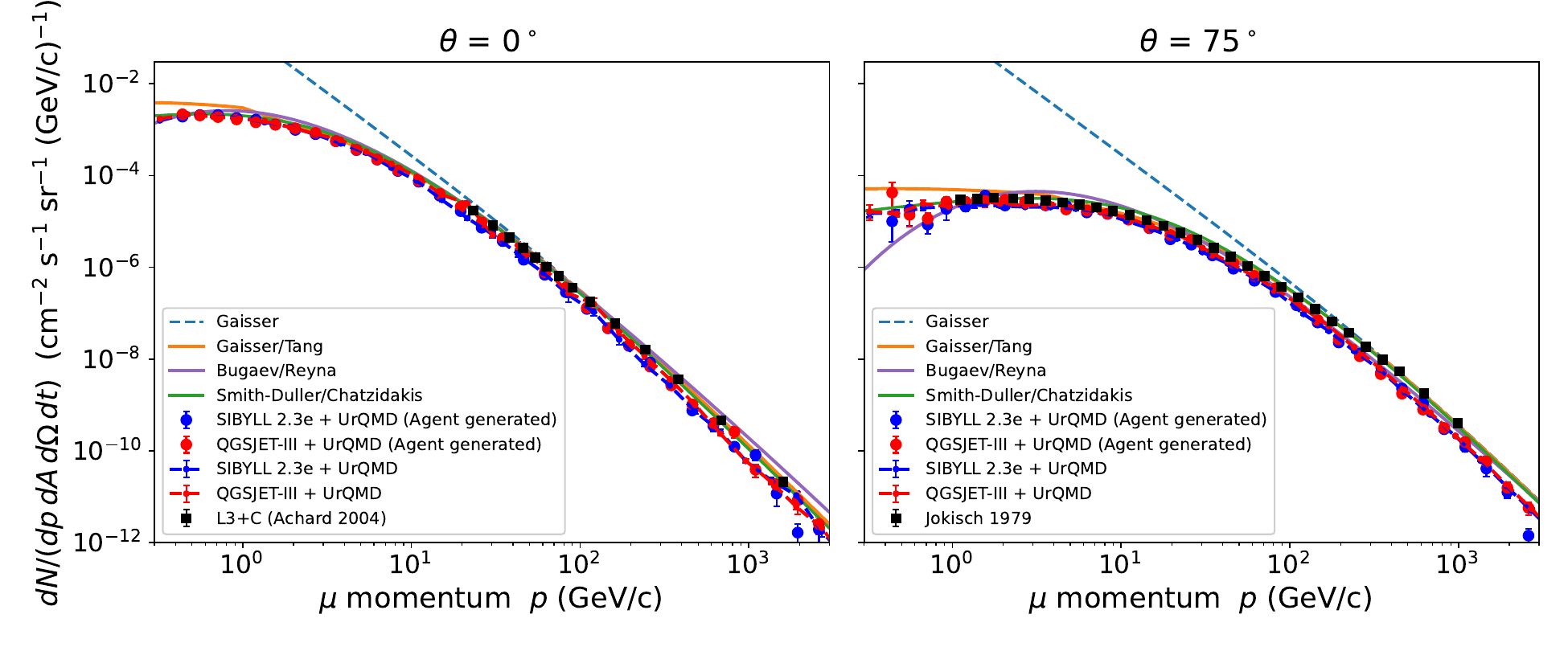}\\[0.8em]
\includegraphics[width=\columnwidth]{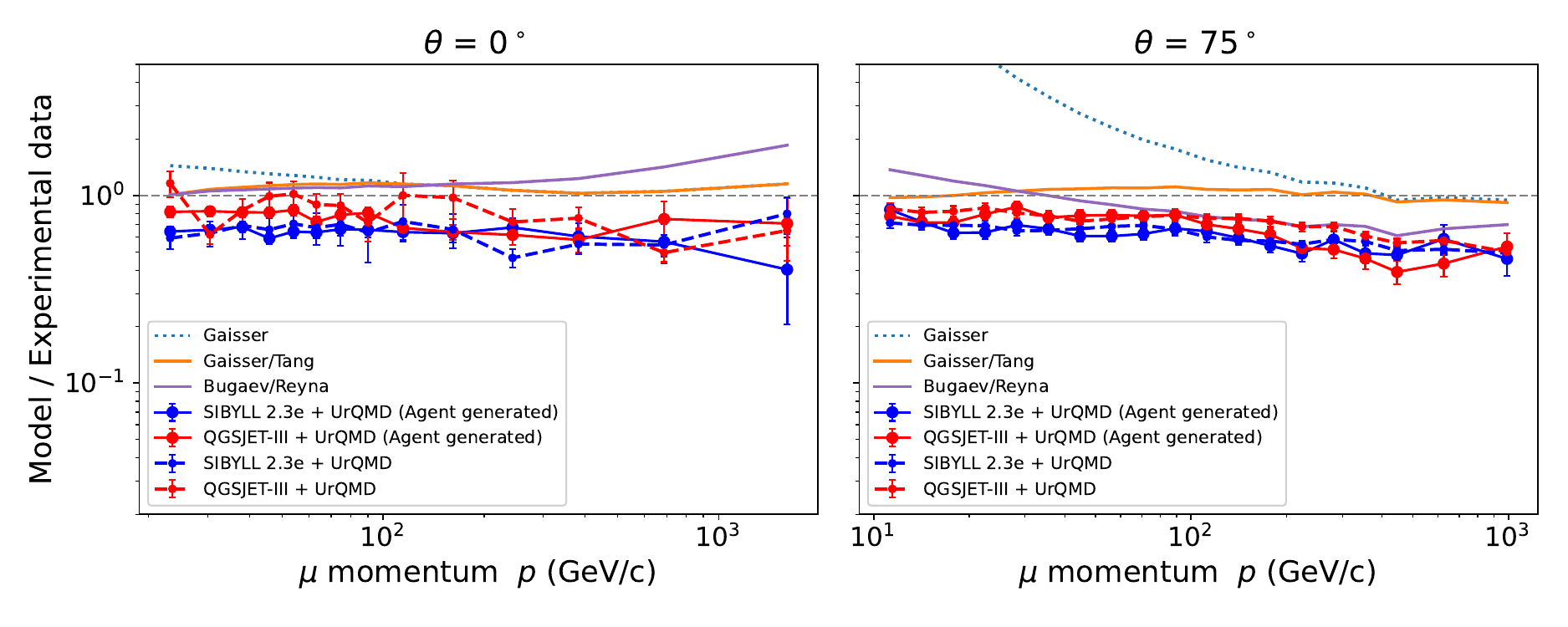}
\caption{Procedural cross-check of the muon-flux calculation.  Top: direct
CORSIKA spectra (curves) and the fixed-energy, fixed-direction results
generated by \textsc{CorsikaAgent} (points), together with analytic references
and the L3+C~\cite{Achard2004} and Jokisch \textit{et al.}
~\cite{Jokisch1979} measurements.  Bottom: the corresponding ratios to data
near $0^\circ$ (left) and $75^\circ$ (right).}
\label{fig:crosscheck}
\end{figure}

This distinction was useful during development.  An intermediate agent
analysis divided a fixed-direction result by a finite angular-bin solid angle,
even though no such angular distribution had been sampled.  The extra factor
raised the predicted intensity by roughly an order of magnitude.  Comparison
with the Gaisser curve and the experimental measurements exposed the error;
inspection of the estimator then identified the double normalization.  The
correction was made before producing the points in Fig.~\ref{fig:crosscheck}.
This example clarifies what the reference calculations are for: validation
tests with known domains of applicability, useful for catching an error but
no substitute for examining the estimator directly.

The upper panels of Fig.~\ref{fig:crosscheck} compare the fixed-energy agent
result with the direct CORSIKA spectra.  The corresponding data ratios are
shown below.  At the graphical precision and statistical reach of these
samples, the agent-generated points follow the direct calculations at both
angles.  A quantitative compatibility figure is deliberately withheld here,
since the present plots lack a complete covariance matrix and the shared
inputs correlate the two predictions.  What the agreement supports is that
the agent faithfully reconstructs the implemented workflow and normalization;
validating CORSIKA or the hadronic models independently would require a
different kind of test.

\begin{figure}[t]
\centering
\includegraphics[width=\columnwidth]{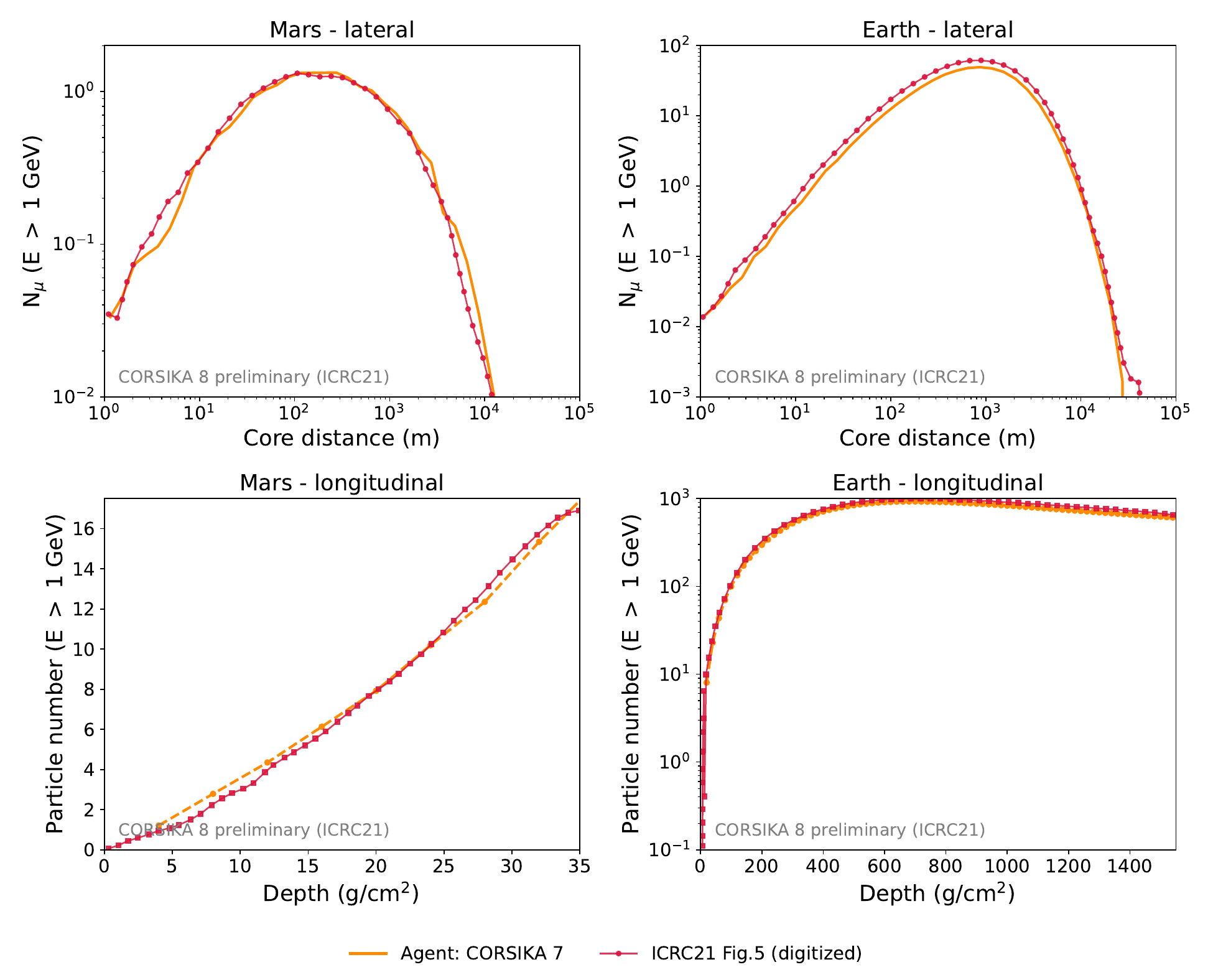}
\caption{Martian-atmosphere benchmark.  Lateral muon densities above
$1\,\mathrm{GeV}$ (top) and longitudinal particle profiles (bottom) are shown
for Mars (left) and Earth (right).  The CORSIKA~7 agent calculation is compared
with curves digitized from the CORSIKA~8 study of Ref.~\cite{Ulrich2021}.}
\label{fig:mars}
\end{figure}

Three additional cases extend the workflow beyond the configuration used for
the sea-level calculation: a transfer test for Mars, a first prediction for
Venus, and an electric-field modification for thunderstorm conditions.  In
the first, the agent
autonomously reconfigures the simulation---replacing the terrestrial
atmosphere with a CO$_2$ profile---and compares a
CORSIKA~7 calculation with digitized CORSIKA~8 results for Mars
~\cite{Ulrich2021}.  Figure~\ref{fig:mars} compares the lateral density of
muons above $1\,\mathrm{GeV}$ and the longitudinal shower development for Mars
and Earth.  Since the reference curve is digitized and the two CORSIKA generations differ, this comparison
speaks to qualitative behavior and shape; a code-to-code identity test would
demand a stricter setup.

\begin{figure}[t]
\centering
\includegraphics[width=\columnwidth]{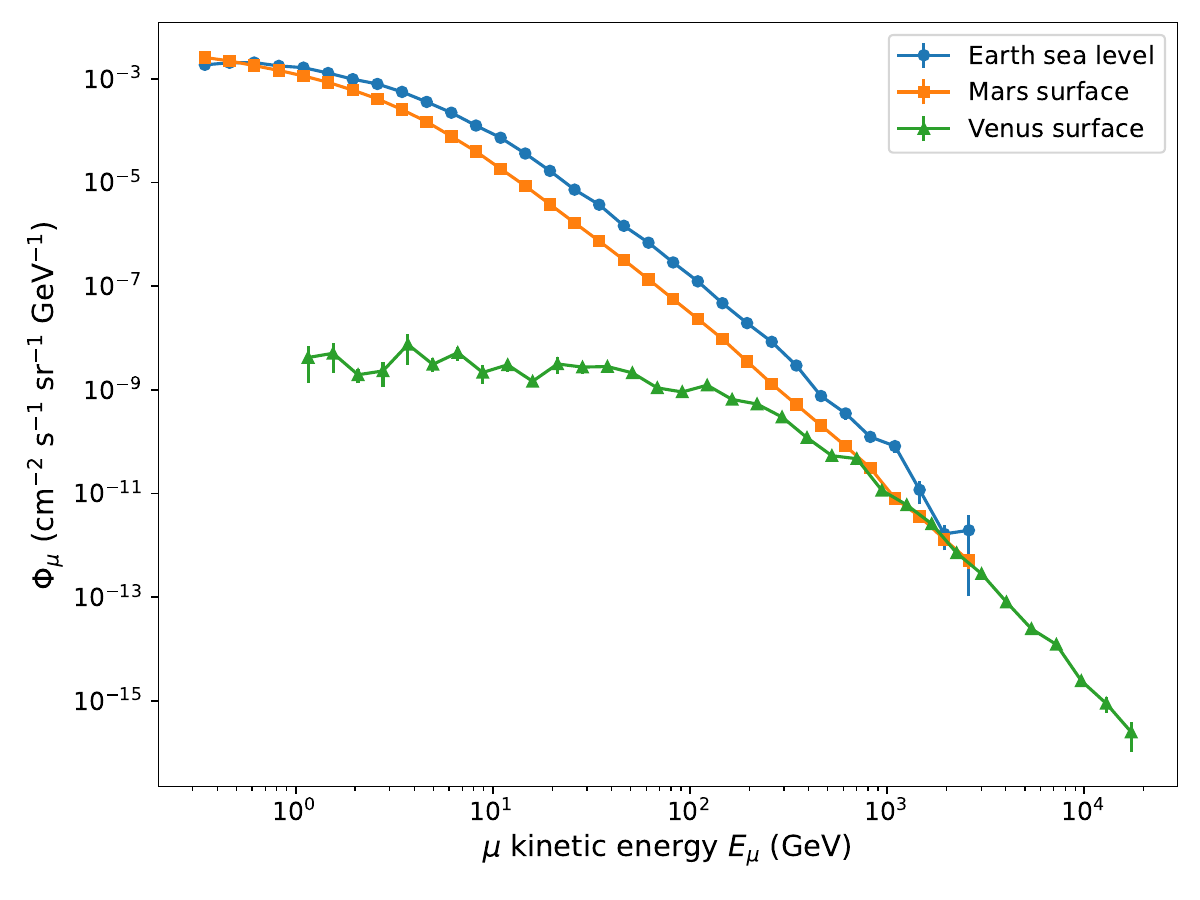}
\caption{Surface muon differential flux obtained with the agent-driven
\textsc{CORSIKA} workflow for the terrestrial sea level, Mars, and Venus, for
proton primaries weighted by the Papini solar-minimum spectrum.  Venus, with
an atmospheric column depth about two orders of magnitude larger than the
Earth's, transmits a muon flux that is orders of magnitude lower than the
terrestrial and Martian fluxes at low energies.}
\label{fig:venus}
\end{figure}

In the second case, the agent is used to produce a first prediction, for
Venus, whose 92-bar
CO$_2$ atmosphere~\cite{NSSDCAVenus} corresponds to a vertical column depth of about
$1.04\times10^5\,\mathrm{g\,cm^{-2}}$---two orders of magnitude larger than
the terrestrial atmosphere---so that the surface flux is suppressed by
orders of magnitude relative to the terrestrial one at low energies.  The
atmospheric profile follows the Venus
International Reference Atmosphere (VIRA) temperature
structure~\cite{Seiff1985}.  Figure~\ref{fig:venus} compares the surface
muon differential fluxes obtained with the same workflow for Venus, Mars, and
the terrestrial sea level.  Earlier calculations of cosmic-ray secondaries
at Venus are confined to the cloud layer: a dose-rate study performed with
the cascade-equation code MCEq~\cite{Globus2021} reports the radiation
environment, and a large-scale AIRES simulation~\cite{Anchordoqui2025}
computes the total particle flux between 40 and 65~km altitude.  No previous
calculation of the surface muon flux at Venus exists, and no measurement has
been made, so this comparison constitutes a falsifiable first
prediction of the workflow rather than a validation.

\begin{figure}[t]
\centering
\includegraphics[width=\columnwidth]{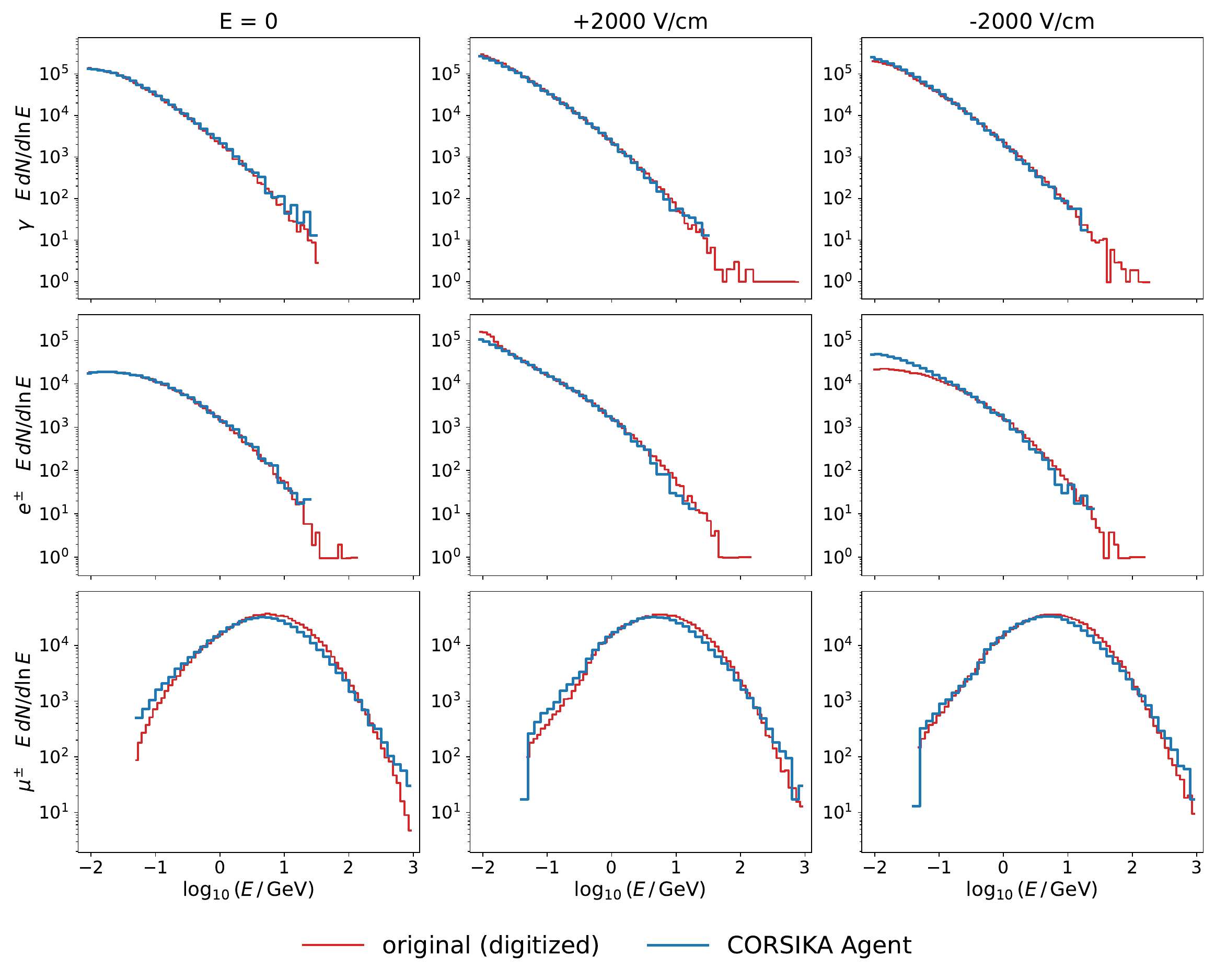}
\caption{Thunderstorm-field benchmark at $1400\,\mathrm{m}$.  The panels show
$E\,dN/d\ln E$ for photons, electrons and positrons, and muons (rows), with
fields of $0$, $+2000$, and $-2000\,\mathrm{V/cm}$ (columns).  Blue curves are
the agent-managed CORSIKA calculation; red curves are digitized from
Ref.~\cite{Abbasi2022}.}
\label{fig:thunder}
\end{figure}

In the third case the agent modifies the steering configuration to add
uniform electric fields of $0$, $+2000$, and
$-2000\,\mathrm{V/cm}$ and records photons, electrons and positrons, and muons
at $1400\,\mathrm{m}$.  These spectra are compared in
Fig.~\ref{fig:thunder} with digitized Telescope Array calculations
~\cite{Abbasi2022}; measurements of thunderstorm-induced muon-rate variations
by GRAPES-3 provide related experimental context~\cite{Hariharan2019}.  The
figure can therefore speak to both the spectral shapes and the response to
field polarity.

Across the three cases, the agent reconstructs a validated result through an
independent sampling path and, in the modified environments, autonomously
reconfigures the simulation---atmosphere, magnetic field, electric field---to
produce quantitative predictions where no measurements exist.  This extends
the workflow from auditable cross-checking toward first-principles
prediction, turning an LLM-driven sequence of tool calls into a reproducible
instrument for atmospheric particle physics on Earth and beyond.

\section{Summary and outlook}\label{sec:summary}

We have calculated the sea-level proton-induced muon flux with CORSIKA~7.8010,
using \textsc{UrQMD} with either \textsc{Sibyll}~2.3e or
\textsc{QGSJET-III}.  After reweighting to the Papini solar-minimum proton
spectrum, both configurations describe the principal momentum and angular
trends in the L3+C and Jokisch data.  Their difference is modest compared with
the full spread of the reference parametrizations shown.  A quantitative
agreement level is not quoted, since many factors affect the comparison and
systematic uncertainties are not considered.

\textsc{CorsikaAgent} reconstructs the same observable with fixed-energy,
fixed-direction samples and reproduces the direct calculation at the precision
resolved here.  The alternative sampling exposed an angular-normalization
error during development.  The Martian-atmosphere and thunderstorm-field cases
demonstrate transfer of the workflow, with their stated limitations drawing a
clear line between shape agreement or a rescaled comparison on one hand and
absolute validation on the other.  Transferred to Venus, the same workflow
produces the first prediction of the surface muon flux, an environment for
which no measurement exists.

Beyond the individual results, the workflow itself argues for wider use.
Its defining advantage is provenance: every steering card, executable, seed,
log, and analysis script is retained together with the number it produced,
so that any flux point can be reproduced or re-examined long after it was
generated.  A second advantage is transferability: the four curated prompts
encode the procedure---atmosphere, gravity, magnetic field, electric
field---once, and the Mars, Venus, and thunderstorm cases show that moving
to a new environment is then largely a matter of exchanging well-documented
parameters rather than rewriting the analysis, with each migration checked
against an independent reference before it is accepted.  The prospects are
correspondingly broad.  Planetary exploration increasingly requires
quantitative radiation environments---for mission planning, surface science,
and potential future muon radiography of planetary interiors---and an
auditable, automatically reconfigurable pipeline is a natural instrument for
supplying them; the same machinery extends to any environment that can be
encoded as an atmosphere, a field configuration, and an observation level.

The dominant physics limitations are the proton-only primary beam, finite
statistics at the largest momenta, and the missing propagation of primary,
atmospheric, and hadronic-model uncertainties.  Addressing those items is
necessary before the calculation can be interpreted as a precision prediction
of the total muon flux.  For the agent, the corresponding next step is to make
these uncertainty variations explicit workflow objects and to evaluate the
system on a larger benchmark set with predefined acceptance criteria.

\begin{acknowledgments}
This work is supported in part by the National Natural Science Foundation of
China under Grant No.~12325504.
\end{acknowledgments}

\section*{Data availability}
The steering cards, analysis scripts, and simulation artifacts supporting
this study are publicly available in the CorsikaAgent repository at
\url{https://github.com/CZ-COLLAB/corsika-agent}.

\bibliographystyle{apsrev4-2}
\bibliography{apssamp}

\end{document}